\documentclass[sigconf]{acmart}
\usepackage{tabularx}
\usepackage{balance}
\AtBeginDocument{%
  }

\copyrightyear{2025}
\acmYear{2025}
\setcopyright{acmlicensed}\acmConference[KDD '25]{Proceedings of the 31st ACM SIGKDD Conference on Knowledge Discovery and Data Mining V.2}{August 3--7, 2025}{Toronto, ON, Canada}
\acmBooktitle{Proceedings of the 31st ACM SIGKDD Conference on Knowledge Discovery and Data Mining V.2 (KDD '25), August 3--7, 2025, Toronto, ON, Canada}
\acmDOI{10.1145/3711896.3737190}
\acmISBN{979-8-4007-1454-2/2025/08}

\begin{document}

\title{An Audio-centric Multi-task Learning Framework for Streaming Ads Targeting on Spotify}

\author{Shivam Verma}
\authornote{Both authors contributed equally to this research.}
\affiliation{%
  \institution{Spotify Inc.}
  \city{London}
  \country{UK}
}
\email{shivamv@spotify.com}

\author{Vivian Chen}
\authornotemark[1]
\affiliation{%
  \institution{Spotify Inc.}
  \city{San Francisco}
  \country{USA}
}
\email{vivianc@spotify.com}

\author{Darren Mei}
\affiliation{%
  \institution{Spotify Inc.}
  \city{New York}
  \country{USA}
}
\email{darrenm@spotify.com}

\renewcommand{\shortauthors}{Shivam Verma, Vivian Chen, and Darren Mei}

\begin{abstract}
Spotify, a large-scale multimedia platform, attracts over 675 million monthly active users who collectively consume millions of hours of music, podcasts, audiobooks, and video content. This diverse content consumption pattern introduces unique challenges for computational advertising, which must effectively integrate a variety of ad modalities—including audio, video, and display—within a single user experience. Traditional ad recommendation models, primarily designed for foregrounded experiences, often struggle to reconcile the platform’s inherent audio-centrality with the demands of optimizing ad performance across multiple formats and modalities. To overcome these challenges, we introduce Cross-modal Adaptive Mixture-of-Experts (CAMoE), a novel framework for optimizing click-through rate (CTR) prediction in both audio-centric and multi-modal settings. CAMoE enhances traditional mixture-of-experts (MoE) models by incorporating modality-aware task grouping, adaptive loss masking, and deep-cross networks (DCN) to capture complex feature interactions within a multi-modal ad ecosystem. Through extensive ablation studies, we demonstrate that this approach achieves near Pareto-optimal performance across audio, video, and display ad formats, significantly improving AUC-PR compared to conventional single-task and content-based multi-task learning baselines. When deployed at scale on Spotify’s ad serving platform, CAMoE delivered substantial gains, yielding a 14.5\% increase in CTR for audio ads, a 1.3\% increase for video ads, and a 4.8\% reduction in expected cost-per-click (eCPC) for audio slots.
\end{abstract}



\maketitle

\section{Introduction} \label{intro}

\begin{figure}[h]
\begin{tabular}{ll}
\includegraphics[scale=0.4]{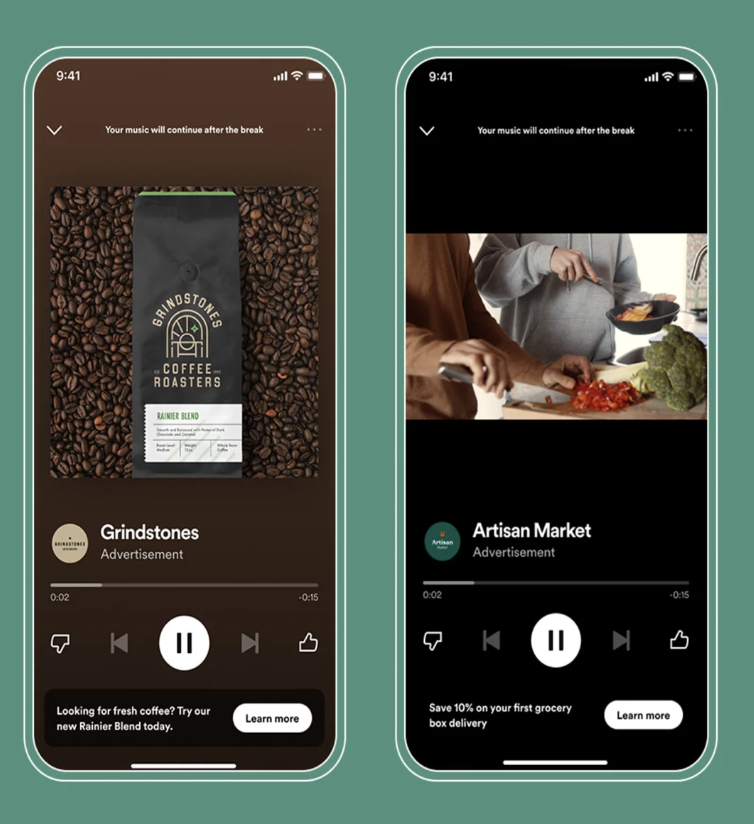}
&
\includegraphics[scale=0.33]{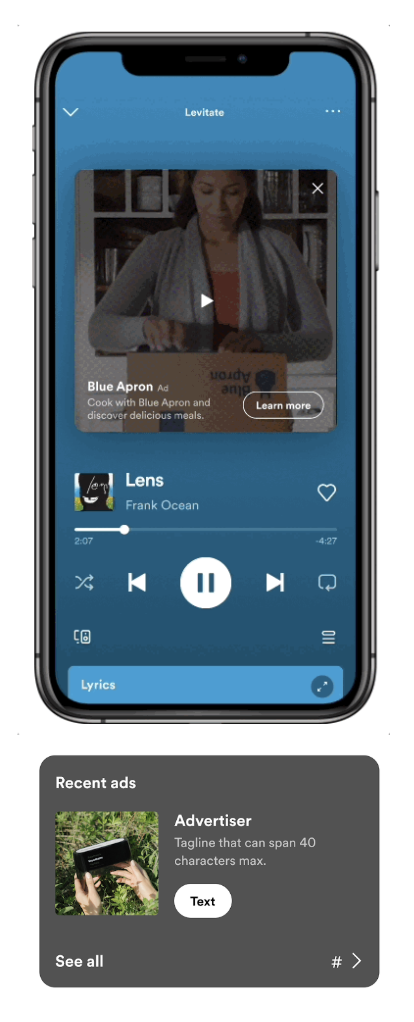}
\end{tabular}
\caption{Left to right: (a) An audio ad stream. (b) An unmuted video ad stream. (c) A muted embedded video ad in the Now Playing View (NPV). (d) A display ad in the NPV. An audio ad typically plays when the user is out-of-focus, while a video ad plays when a user is in-focus.}
\label{fig:fig1}
\end{figure}

Spotify, one of the world's largest audio streaming platforms, attracts over 675 million monthly active users. A significant portion of Spotify's revenue comes from advertising, making accurate ad click-through rate (CTR) prediction critical for both revenue optimization and user experience. Unlike many visually-dominated online platforms, Spotify is \emph{audio-centric}, with users primarily engaging with audio content, often in a backgrounded state. This presents a unique and commercially important challenge for ad targeting: how to effectively serve a mix of audio, video, and display ads (see Fig. \ref{fig:fig1}) within a predominantly audio-driven user experience.

Existing ad targeting models, typically designed for visually-rich platforms, are not well-suited to this audio-centric, multi-modal environment. Single-task models, trained separately for each ad format, suffer from data sparsity and increased operational complexity. While multi-task learning (MTL) offers potential benefits by sharing information across tasks, standard MTL approaches often fail to account for the significant differences in user engagement and CTR distributions across modalities. Furthermore, common techniques for handling data imbalance prove insufficient for the extreme skew present on Spotify.

In this paper, we present \textbf{Cross-modal Adaptive Mixture-of-Experts (CAMoE)}, a novel multi-task learning framework \emph{specifically designed and deployed} to address the challenges of CTR prediction on Spotify's audio-centric, multi-modal ad platform. CAMoE introduces three key innovations:

\begin{enumerate}
    \item \textbf{Modality-Aware Task Grouping:} We partition the CTR prediction task into distinct heads based on ad \emph{modality} (audio vs. video), allowing the model to learn specialized representations that capture the unique user behaviors associated with each format. 
    \item \textbf{Adaptive Loss Masking (ALM):} To mitigate the severe data imbalance, we introduce ALM, a novel adaptation of loss masking techniques that restricts each task's updates during training to examples from the same modality. This prevents the dominant audio task from overwhelming the learning of other tasks and improves calibration.
    \item \textbf{Deep \& Cross Network (DCN) Experts:} We integrate DCN within each expert of the Mixture-of-Experts (MoE) architecture. This leverages DCN's strength in capturing complex feature interactions \emph{within} each modality, enhancing the model's ability to learn nuanced patterns.
\end{enumerate}

The rest of the paper is organized as follows: In Section \ref{relatedwork}, we review related work in multi-task learning, Mixture-of-Experts architectures, feature interaction learning, and handling data imbalance in recommender systems. Section \ref{problemsetting} details the problem setting, including a description of Spotify's ad platform and the challenges of multi-modal advertising. Section 4 describes the multi-task setup, including task definitions. In Section \ref{architecture}, we present the CAMoE architecture and its key components. Section \ref{exp} details our experimental methodology, including offline evaluations, ablation studies, and the results of a large-scale online A/B test on Spotify's ad serving platform. Finally, Section \ref{conclusion} presents our conclusions. These results underscore the effectiveness of modality-aware architectures in real-world, multi-modal advertising systems, \emph{offering a practical, deployed solution to a commercially critical challenge in the rapidly growing audio advertising space.}

\section{Related Work} \label{relatedwork}

Our work builds upon and extends prior research in several key areas: multi-task learning (MTL), Mixture-of-Experts (MoE) architectures, feature interaction learning, handling imbalance in recommender systems, and multi-modality.

\textbf{Multi-Task Learning in Recommender Systems.} MTL aims to improve performance by learning multiple related tasks simultaneously \cite{Caruana1997}. It has been successfully applied in recommender systems for tasks like predicting user engagement \cite{Covington2016, Zhao2019, airbnb2023, googleads2022, lirank2024}, candidate generation \cite{wang2023}, and optimizing multiple objectives \cite{Lin2019, Long2017, Sener2018}. While many MTL approaches focus on what tasks to learn together \cite{Standley2020}, our work emphasizes how to group tasks based on modality in an audio-centric, multi-modal environment. We also differ by explicitly addressing the significant data imbalance inherent in this setting, a problem less emphasized in general MTL recommenders.

\textbf{Mixture-of-Experts Architectures.} MoE architectures enhance model capacity and efficiency by dynamically routing inputs to specialized "expert" sub-networks \cite{shazeer2017}. Multi-gate Mixture-of-Experts (MMoE) \cite{Ma2018} extends this by sharing experts across tasks while using task-specific gating. Our CAMoE framework builds on MMoE but introduces modality-aware task grouping, which differs from prior work that primarily focuses on sharing experts across all tasks or using content-based groupings. Recent work has also looked at modality-aware MoE \cite{lin2024}, but in the context of large language models, not recommender systems with explicit interaction learning.

\textbf{Feature Interaction Learning.} Capturing feature interactions is crucial for accurate CTR prediction in advertising \cite{McMahan2013, wang2020, wang2017, wang2021}. Wide \& Deep networks \cite{wdl2016} combine linear models (memorization) with deep networks (generalization). Deep \& Cross Networks (DCN) \cite{wang2017, wang2021, li2024dcnv3} explicitly model feature crosses at each layer. Our work integrates DCN within each expert of the CAMoE framework, leveraging its ability to capture complex interactions within each modality-specific task. This differs from simply using DCN as a standalone model, as our approach combines the benefits of feature interaction learning with the modality specialization of the MoE architecture.

\textbf{Handling Imbalance in Recommender Systems.} Data imbalance is a common challenge in recommender systems, particularly in advertising where positive examples (e.g., clicks) are often rare \cite{Ma2018}. Common techniques include oversampling, undersampling, and cost-sensitive learning, and focal loss \cite{lin2017focal}. Ma et al. \cite{esmMa2018} touch on imbalance by considering entire-space modeling. 

\textbf{Multi-modality in Recommender Systems.} Multi-modal recommender systems leverage information from different modalities (e.g., text, images, audio, video) to improve recommendations \cite{Wei_2019_ACMMM}. Previous works have looked into multimodal recommendations \cite{zhang2020look}, but ours is to our knowledge, the first to specifically tackle audio-centric platforms. 

\section{Problem Setting} \label{problemsetting}

\begin{table*}[t]
  \caption{Multi-modal Ad Placements (Slots) on Spotify}
  \label{tab:slot_distribution}
  \begin{tabular}{l l p{12cm}}
    \toprule
    \textbf{Ad Slot} & \textbf{Modality} & \textbf{Description} \\
    \midrule
    Stream Audio & \textit{Audio} & Serves an in-stream audio ad in a music context. \\ 
    Podcast & \textit{Audio} & Serves an in-stream audio ad in a podcast context \\
    Stream Video & \textit{Video} & Serves an unmuted in-stream video ad in a music context \\
    Embedded Music & \textit{Video} & Serves a muted Now Playing View (NPV) video ad in a music context \\
    Podcast Video & \textit{Video} & Serves an unmuted in-stream video ad in a podcast context \\
    Stream Audio Leavebehind & \textit{Display} & Serves a display click-through action (CTA) card in a music context accompanying an already streamed audio ad. Not counted as an additional impression \\
    Podcast Leavebehind & \textit{Display} & Similar to Stream Audio Leavebehind, display CTA card in a podcast context \\
    \bottomrule
  \end{tabular}
\end{table*}

\subsection{Audio-Centric Platform} \label{audiocentric}

Spotify, while being a multi-modal platform offering both video and audio content, is fundamentally different from many online advertising environments due to its audio-centric nature. The user experience is primarily driven by audio content consumption (music, podcasts, audiobooks), leading to distinct user behaviors and engagement patterns that significantly impact ad targeting. A crucial aspect of user behavior on Spotify is the concept of \emph{focus state}.

We define a user as being \emph{in-focus} when the Spotify application is actively displayed on their device's screen. Conversely, a user is \emph{out-of-focus} when the app is running in the background, or the device's screen is off. This distinction is critical because it directly impacts user attention and, consequently, click-through rates (CTRs).

Several key challenges arise from Spotify's audio-centric nature:
\begin{itemize}
\item \textbf{Predominantly Out-of-Focus Consumption:} Over 70\% of user listening time on Spotify occurs while the user is out-of-focus. This is in stark contrast to visually-driven platforms, where users are typically in-focus.
\item \textbf{Significant CTR Disparity between Focus States:} The difference in user attention between in-focus and out-of-focus states translates to a significant disparity in ad CTRs. Internal analysis reveals that in-focus impressions exhibit CTRs that are, on average, ten times higher than those of out-of-focus impressions. This highlights the premium value of in-focus impressions.
\item \textbf{Highly Skewed Ad Inventory:} Audio ad impressions constitute the vast majority of the total ad inventory on Spotify. This creates a severe data imbalance, making it difficult for standard models to effectively learn from the relatively scarce, but higher-value, video ad impressions.
\item \textbf{Implicit Interaction Signals:} The nature of interacting with audio ads, often while multi-tasking or with the screen off, introduces a significant element of implicitness to user signals, such as click. This contrasts with more explicit interactions typical of visual ad formats.
\end{itemize}

These characteristics create a challenging environment for traditional ad targeting models. Models trained on this data tend to be heavily biased towards the dominant audio modality, leading to suboptimal performance and miscalibration on video ads. This necessitates a fundamentally different approach to ad targeting, one that explicitly accounts for the unique characteristics of an audio-centric, multi-modal platform.

\begin{figure}[h]
\includegraphics[scale=0.3]{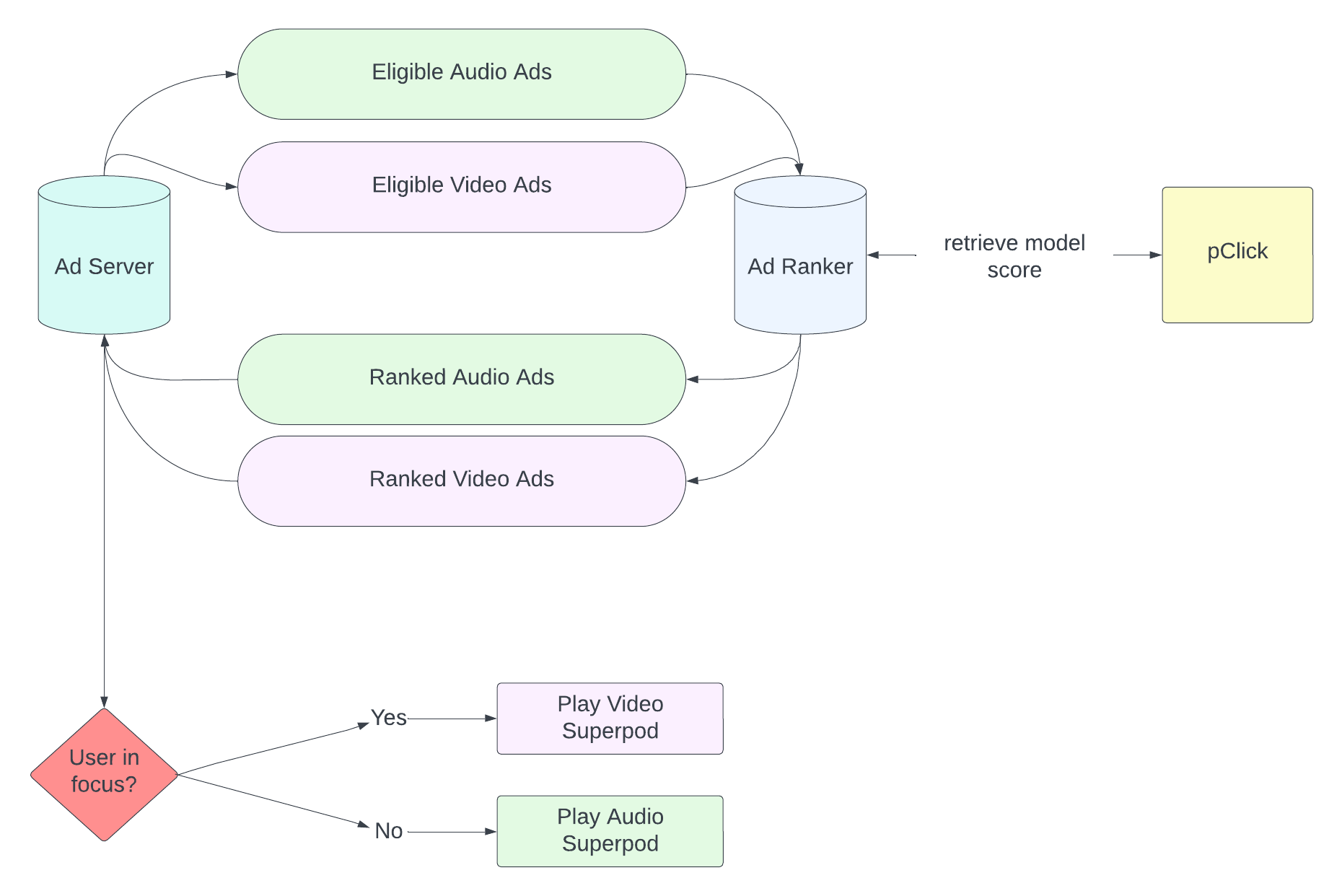}
\caption{High-level overview of the ads ranking system at Spotify.}
\label{fig:adserver}
\end{figure}

\subsection{Ad Ranking System} \label{adranker}

Fig. \ref{fig:adserver} illustrates Spotify's multi-stage ad ranking system, which selects an ad for each user based on a combination of campaign objectives (clicks, reach, impressions), budget constraints, pacing, and predicted user engagement. The process begins with a pool of eligible ads that meet essential targeting criteria (e.g., demographics, location, listening history) and categorizes them into audio and video modalities.

At the core of the system is the \textit{Ad Ranker}, which assigns each eligible ad a predicted click probability (pCTR) generated by a machine learning model. This score is then used to compute a Final Bid for each ad, particularly for click-optimized campaigns, as follows:

\begin{equation} \label{bideqn}
\text{Final Bid} = \min\left(\left\lceil\frac{o}{1 + p} \cdot \frac{c}{b}\right\rceil, o\right)
\end{equation}
\begin{math}
\begin{aligned}
\text{where \ }
c &\text{ = pCTR,} \\
b &\text{ = Average CTR for ad over past 24 hours,} \\
p &\text{ = Pacing Multiplier,} \\
o &\text{ = Max Bid.} \\
\end{aligned}
\end{math}
\\
Ads are then ranked based on their \textit{Final Bid} in a generalized second-price auction. The ranked ads are subsequently organized into separate lists for audio and video formats, with the system tailoring the ad presentation to the user’s context: a Video ad pod (\textit{superpod}) is served when the user is in focus, whereas an Audio ad pod is served when the user is out of focus.

\begin{figure*}[h]
\includegraphics[scale=0.3]{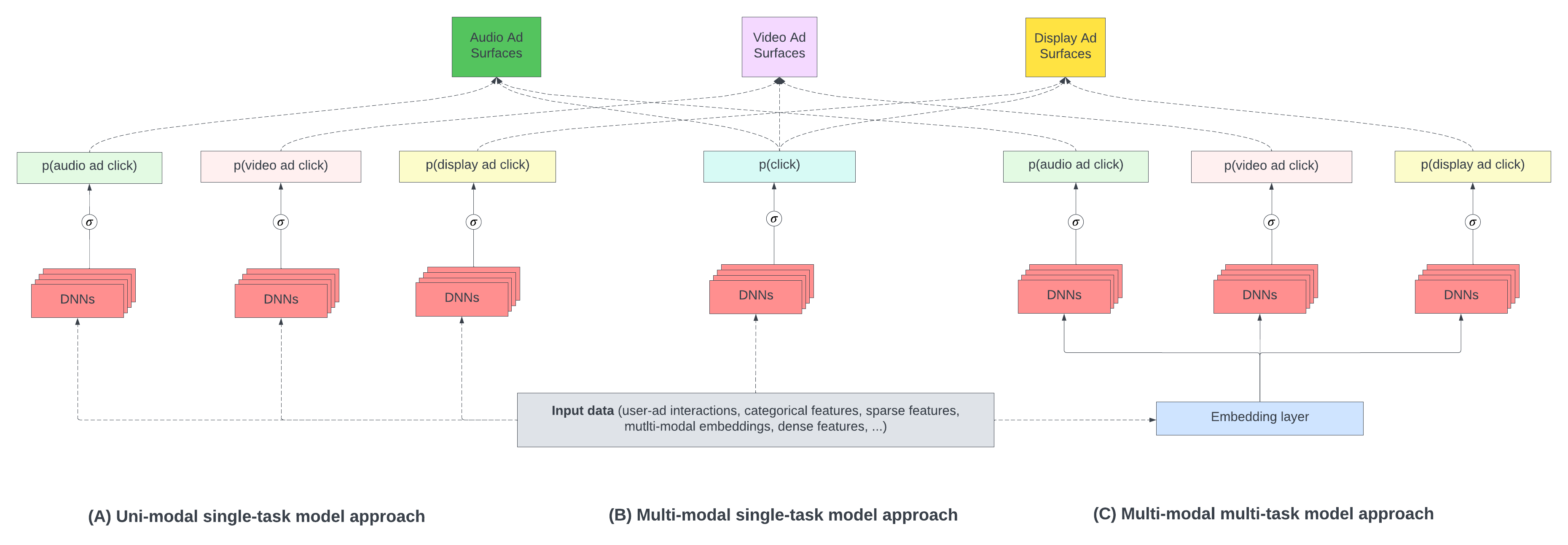}
\caption{(A) Independent uni-modal pClick models, for each ad surface or modality (B) Single-task pClick model incorporating multiple modalities or ad surfaces. (C) Multi-modal multi-task pClick model, with heads predicting outcomes for various modalities - audio, video and display ads.}
\label{fig:mtlabc}
\end{figure*}

\subsection{Addressing Multiple Ad Modalities}

Spotify's predominantly out-of-focus listening, the significant CTR disparity between in-focus and out-of-focus impressions, the highly skewed ad inventory towards audio, and more implicit interaction signals, all contribute to the difficulty of accurately predicting click-through rates across its diverse ad modalities (audio, video, and display). Before developing CAMoE, we explored two initial approaches to address this multi-modal challenge:
\begin{enumerate}
    \item \textbf{Uni-modal Single-Task Models:} This involved training separate, independent \texttt{pClick} models (Fig. \ref{fig:mtlabc}A) for each ad slot (e.g., \textit{Stream Audio}, \textit{Podcast Video}, \textit{Embedded Music}). While this allows each model to specialize in a particular ad format, it suffers from data sparsity for less frequent slots (like \textit{Podcast Video}) and significantly increases operational overhead as new ad formats are introduced.
    \item \textbf{Multi-modal Single-Task Model:} This approach uses a single \texttt{pClick} model trained across \emph{all} ad slots (Fig. \ref{fig:mtlabc}B). By leveraging all available data, this aims to improve generalization. However, due to the dominance of audio impressions and the differences in user engagement, this single model tends to be biased towards the audio modality, leading to suboptimal performance and miscalibration, particularly for video ad slots.

\end{enumerate}

Neither of these initial approaches adequately addressed the core challenges of Spotify's advertising ecosystem. The uni-modal approach was hampered by data limitations, and the multi-modal single-task model struggled to balance the diverse needs of different ad modalities within a single, unified model. These challenges motivated the development of an MTL framework (Fig. \ref{fig:mtlabc}C) that overcomes these limitations through modality-aware optimizations.

\section{Multi-task Setup} \label{mtlsetup}

\subsection{Dividing the Click Task by Modality}
Designing the multitask model posed a significant challenge in balancing the granularity of tasks with their inherent similarities and shared characteristics. To address this, we partitioned the click prediction task into two distinct heads — one for audio ad slots and one for video ad slots. This approach was motivated by the substantial differences in user engagement and the magnitude of CTR across formats.

The audio task encompasses not only audio formats, such as \textit{Stream Audio} and \textit{Podcast} ads, but also integrates display formats - \textit{Stream Audio Leavebehind} and \textit{Podcast Leavebehind} slots — since they are not counted as independent impressions. In contrast, the video task is dedicated to ad formats that inherently drive higher CTRs due to their visual and actively engaging nature, such as \textit{Stream Video}, \textit{Embedded Music}, and \textit{Podcast Video} ads. See Table \ref{tab:slot_distribution} for more information about the differences between these slots. This separation also allows for the implementation of independent calibration layers (see Subsection \ref{calibsec}), where the video head focuses on refining predictions in scenarios with higher click probabilities, and the audio head is tuned to better calibrate the sparser click distributions observed in audio-driven contexts.

\subsection{Data Imbalance} \label{dataimb}
Our click dataset exhibits a significant class imbalance, with the \textit{Stream Audio} slot overwhelmingly dominating, while other slots, particularly \textit{Podcast Video} and \textit{Embedded Music}, suffer from limited data availability, hindering generalization. A naive model trained on this imbalanced data will be heavily biased towards the prevalent \textit{Stream Audio} slot, leading to poor performance and miscalibration on minority slots. To mitigate this issue, we explored and evaluated several techniques, including up- or down-sampling minority/majority slots, cost-sensitive learning using class weights, and focal loss function \cite{lin2017focal} that dynamically scales the cross-entropy loss, down-weighting the contribution of well-classified examples. In practice, we found that a combination of down-sampling and adaptive loss masking provided the best balance between overall performance and calibration, particularly for the minority slots.

\subsection{1 vs. 2 vs. 7 tasks} \label{127tasks}
A single-task model that combines all slots fails to capture the intrinsic differences between audio and video modalities. Although a 1-task model achieves overall performance comparable to a 2-task model, it forces the model to learn a compromise representation, leading to significant calibration errors — particularly for low-occurrence slots like podcasts. This indicates that while the model can differentiate clicks from non-clicks on a global scale, its compromised representation results in miscalibration that is exacerbated for slots with limited data exposure.

Conversely, a task-per-slot setup enables maximum specialization but introduces other challenges. Low-occurrence slots, such as \textit{Podcast Video} and \textit{Embedded Music}, suffer from overfitting due to insufficient training data. Furthermore, increasing the number of tasks unnecessarily inflates model complexity and computational overhead without yielding significant performance improvements. This approach also misses opportunities for shared learning between slots that exhibit similar user behaviors, such as those within the audio domain in Subsection \ref{offeval}, we evaluate performance across various such multi-task configurations.

\subsection{Splitting by Content Type} \label{splitcontent}

Beyond format-based grouping (audio vs. video), we explored alternative ways to cluster ad slots. One approach considers content type, where music-related slots (e.g., \textit{Stream Audio}, \textit{Embedded Music}, \textit{Stream Video}) form one group, and podcast-related slots form another.

The distinction between music and podcasts may be significant due to differences in user engagement patterns — for instance, podcast listeners may exhibit more active listening behavior, where audio remains the primary focus. We evaluate the performance between our proposed audio / video modality task setup versus a music / podcast content task setup in Subsection \ref{offeval}.

\section{Architecture} \label{architecture}
\begin{figure}[h]
\includegraphics[scale=0.38]{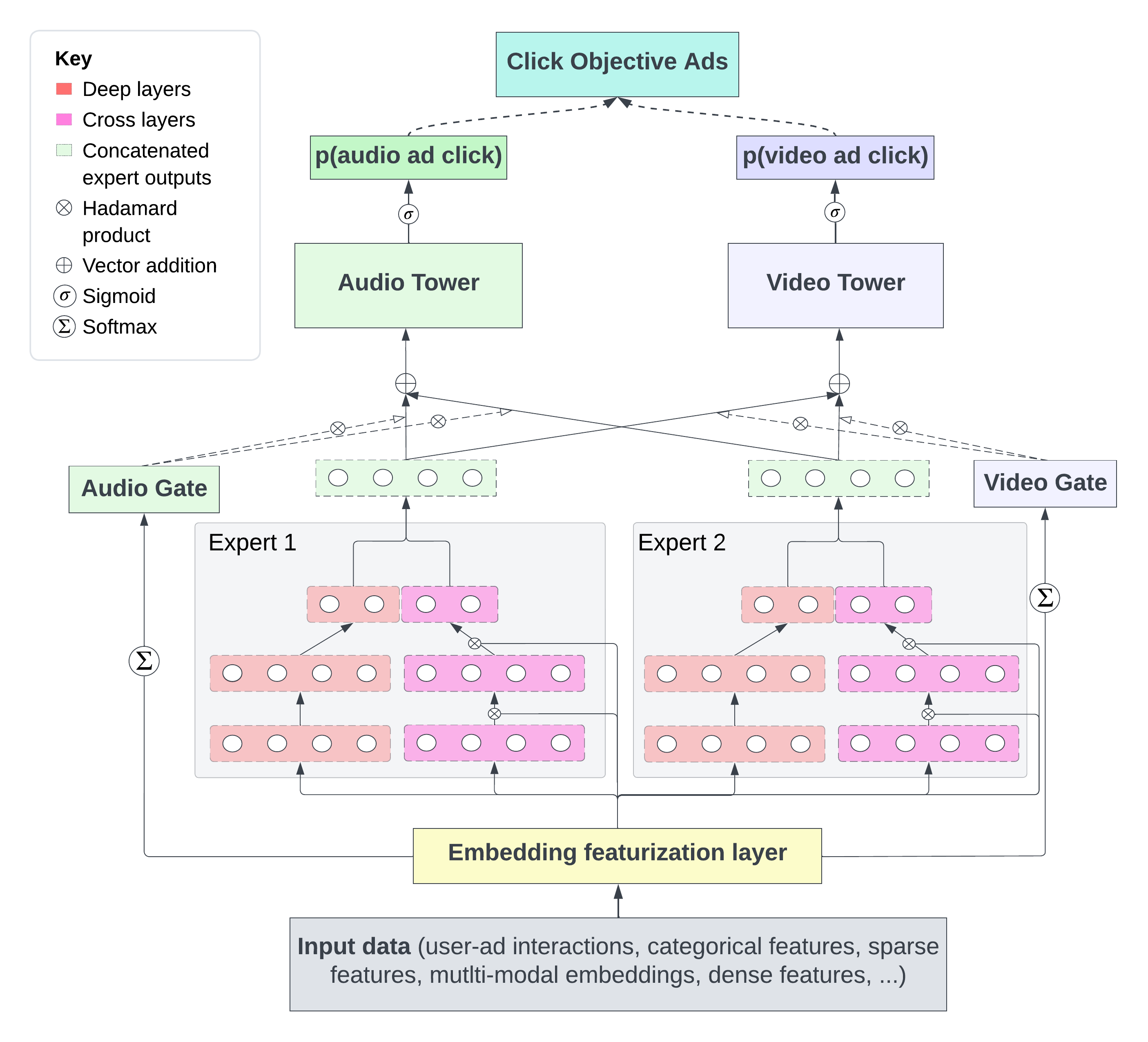}
\caption{Cross-modal Adaptive Mixture-of-Experts (CAMoE) architecture for a 2-task audio-video setup.}
\label{fig:camoe}
\end{figure}

We employ a modified Multi-gate Mixture-of-Experts (MMoE) architecture \cite{Ma2018} (see Fig. \ref{fig:camoe}) for click-through rate prediction. The model features a shared bottom featurization layer with batch normalization, combined with task-specific gating mechanisms that allocate feature representations to a set of experts. Each expert is composed of ReLU-activated DCN-v2 blocks — low-rank deep-cross networks \cite{wang2021} — that capture higher-order feature interactions. Task-specific gates apply a \textit{softmax} activation to the embedding layer, and are then multiplied element-wise with the concatenated expert outputs to obtain the final inputs to the task-specific towers.

\subsection{Adaptive Loss Masking} \label{alm}
We use an adaptively-masked binary cross-entropy loss that restricts each task's updates to events from the same modality. For instance, the audio click task updates solely based on clicks or impressions attributed to an audio or display slot. For a $M$-task model with task modalities $m \in M$, let $\mathcal{\tilde{L}}_m^{(i)}$ denote the contribution of example $i$ with modality $m$ to the overall loss.

\begin{multline}
\mathcal{\tilde{L}}_m^{(i)} = \sum_{n \in M} \mathbb{I}\{ n = m\}\Big[ y_i^{(n)} \log \hat{y}_i^{(m)} + \\ (1 - y_i^{(n)}) \log (1 - \hat{y}_i^{(m)}) \Big]
\end{multline}
Then the adaptive loss-masked $M$-task loss function is defined as
\begin{equation}
\mathcal{L}_M(y, \hat{y}) = -\frac{1}{N} \sum_{i=1}^{N}\sum_{m \in M}\lambda_{m}\  \mathcal{\tilde{L}}_{m}^{(i)}(y_i, \hat{y_i}) 
\end{equation}
where:
\begin{itemize}
  \item $\lambda_{m}$ is a scalar hyperparameter that scales the loss value for modality $m$, s.t. $\sum_{m\in M} \lambda_m = 1$.
  \item $\mathbb{I}\{ \cdot \}$ is the indicator function, which equals 1 if the condition is true and 0 otherwise.
\end{itemize}

Without adaptive masking, each example's loss would backpropagate through both networks, untying the modal awareness of our task-specific towers. As we demonstrate in the results, this approach helps with modal-imbalanced distributions and improves calibration error. In Subsection \ref{appendix:expertstudy}, we also discuss an inference-time expert-masking implementation for forced expert specialization.

\subsection{Deep Cross Networks (DCN)} \label{dcn}
Deep-cross networks \cite{wang2017, wang2021} are a popular architectural choice for recommender systems for ad click predictions - explicitly designed to capture feature interactions. We employ a DCNv2-based expert architecture, consisting of 3 parallel \textit{deep} and \textit{cross} networks. Our choice was determined by a requirement for an expert architecture that could efficiently handle complex interactions between audio and video attributes.

For each expert \( k \) in the mixture-of-experts model, the DCN-v2 block applies a series of \( L \) cross layers to the input feature vector \( \mathbf{x}_0 \). At the \( l \)th layer (\( l = 0,1,\ldots,L-1 \)), the update is defined as:
\begin{equation} \label{dcneqn}
\mathbf{x}_{l+1}^{(k)} = \mathbf{x}_l^{(k)} + \mathbf{x}_0 \odot \left( \mathbf{W}_l^{(2,k)}\, \sigma\Big( \mathbf{W}_l^{(1,k)}\, \mathbf{x}_l^{(k)} + \mathbf{b}_l^{(k)} \Big) \right)
\end{equation}
where:
\begin{itemize}
    \item \( \mathbf{x}_0 \in \mathbb{R}^d \) is the original input feature vector.
    \item \( \mathbf{x}_l^{(k)} \in \mathbb{R}^d \) is the output of the \( l \)th cross layer for expert \( k \).
    \item \( \mathbf{W}_l^{(1,k)} \in \mathbb{R}^{r \times d} \) is the first weight matrix for the \( l \)th layer of expert \( k \).
    \item \( \mathbf{W}_l^{(2,k)} \in \mathbb{R}^{d \times r} \) is the second weight matrix for the \( l \)th layer of expert \( k \).
    \item \( \mathbf{b}_l^{(k)} \in \mathbb{R}^{r} \) is the bias vector for the \( l \)th layer.
    \item \( \sigma(\cdot) \) is an activation function (e.g., ReLU).
    \item \( \odot \) denotes element-wise (Hadamard) multiplication.
\end{itemize}

\subsection{Calibration} \label{calibsec}
In an auction-based ad ranking system, accurate probability calibration is crucial. Miscalibrated predictions can lead to suboptimal ad ordering, revenue loss, and pricing misalignment. While metrics like AUC-PR measure ranking performance, they do not guarantee well-calibrated probabilities.
A well-calibrated model produces predicted probabilities that align with observed click frequencies. Calibration plots (Fig. \ref{fig:masking_calib} \& \ref{fig:dcn_calib}) visualize this relationship. Deviations from the diagonal line indicate miscalibration.
We employ post-training calibration via per-task temperature scaling \cite{calibnn2017}. This method, a simplified version of Platt scaling, introduces a single temperature parameter \textit{T} per task to rescale the model's logits before the sigmoid function. We chose temperature scaling over Platt scaling or isotonic regression due to its simplicity and computational efficiency, as it requires only a single parameter to be tuned per task. Furthermore, our experiments indicated that temperature scaling provided comparable calibration performance to more complex methods, while being significantly faster to train and deploy.

\begin{equation}
P(y = 1 \mid \mathbf{x}) = \frac{1}{1 + \exp(-z / T)}
\end{equation}
where:
\begin{itemize}
    \item \( z \) is the logit for the positive class,
    \item \( T \) is the temperature parameter (\( T > 0 \)).
\end{itemize}

\section{Experiments \& Results} \label{exp}

\subsection{Evaluation Metrics} \label{eval}

The primary metric used in our offline evaluation is AUC-PR, or area under the precision-recall curve, which quantifies how well a model can distinguish between classes. This metric was chosen because it handles class imbalance better than AUC-ROC and binary accuracy, while still taking into account the full range of model predictions. Additionally, Expected Calibration Error (ECE) is used to evaluate calibration. ECE is computed by comparing the accuracy against the model's confidence across a range of bins determined by the prediction distribution \cite{calibnn2017}.

\begin{equation}
\text{ECE} = \sum_{m=1}^{M} \frac{|B_m|}{n} \left| \text{acc}(B_m) - \text{conf}(B_m) \right|
\end{equation}

An important evaluation consideration is that \textit{Stream Audio} and \textit{Stream Video} are both the most frequent and highest-value slots, representing Spotify’s primary ad formats. Therefore, improvements in these slots are prioritized over others in our experiments. This relates to the discussion on near-Pareto optimality in Subsection \ref{appendix:pareto}. 

\subsection{Offline Evaluation Results} \label{offeval}
We compared CAMoE against our production baseline, a Wide \& Deep \cite{wdl2016} network single-task model (as depicted in Fig. \ref{fig:mtlabc}B). We also add comparisons with a uni-modal single-task optimized for our most prominent ad slots (see Fig. \ref{fig:mtlabc}A).

\subsubsection{1 vs. 2 vs. 7 Tasks}
Table \ref{tab:mtl_comparison} presents a baseline comparison across 1-, 2- and 7-task configurations of the CAMoE pCTR setup, with the key difference being the grouping of ad slots into tasks. These results shed light on the trade-offs between different levels of task granularity (simpler to more complex) and their impact on the performance of high and low frequency ad slots.

The 2-task CAMoE, with \textit{audio} and \textit{video} heads, demonstrated the largest overall improvement in AUC-PR across all key formats - \textit{Stream Video} (18.03\%), \textit{Stream Audio} (16.10\%), and \textit{Podcast} (79.62\%) - outperforming the respective 1-task and 7-task configurations. Results for the 1-task setup demonstrate improvements in AUC-PR for the two largest ad slots - \textit{Stream Audio} and \textit{Stream Video} - but to a lesser extent than the 2-task CAMoE. Notably, the \textit{Stream Audio Leavebehind} performed the best (28.99\%) for this setup. However, while AUC-PR improved with this task, we observed that the calibration error worsened significantly.

The 7-task configuration, where each ad slot is treated as a separate head, exhibited the highest AUC-PR gain for \textit{Embedded Music} (88.03\%) among all configurations, suggesting that task specializing does allow the model to capture nuanced slot-specific patterns. However, performance on core slots (\textit{Stream Audio} and \textit{Stream Video}) was notably lower compared to the 2-task model. This implies that excessive fragmentation may hinder generalization, especially for slots that benefit from shared feature representations.

The inability to balance the needs of the more dominant and rare slots proved to be a limitation in the polarized setups of the 1-task and 7-task models. Therefore, splitting the click task into two heads proved to be the optimal task grouping (see Subsection \ref{appendix:pareto}).

\subsubsection{Splitting the Click Task by Content Type}
We also explored slot bundling by content type, segmenting ad slots into a \textit{music} head and a \textit{podcast} head. This seemed to be a promising alternative approach to task grouping, allowing the model to learn from content-specific ad interactions. The music head included \textit{Stream Audio}, \textit{Stream Audio Leavebehind}, and \textit{Embedded Music} slots, while the podcast head contained \textit{Podcast}, \textit{Podcast Video}, and \textit{Podcast Leavebehind} slots. This setup aimed to leverage distinctions in user engagement based on content consumption.

As shown in Table \ref{tab:table3}, this configuration improved performance for \textit{Podcast} and \textit{Podcast Leavebehind} slots, but gains were less pronounced for \textit{Podcast Video}. However, music-aligned slots performed worse compared to modality-based task groupings. 

This approach proved less effective than an audio-video task setup. One possible explanation is that content type alone does not fully capture differences in user interaction, whereas modality-based grouping better aligns with ad engagement patterns.

\begin{table*}[t]
    \centering
    \caption{Comparison of AUC-PR across different single and multi-task models (with \textit{t} tasks). Results are reported as \% change in AUC-PR relative to the Wide\&Deep baseline. Results are statistically significant (p-value < 0.05).}
    \label{tab:mtl_comparison}
    \begin{tabular}{lccccccc}
        \toprule
        \textbf{Model} & {\small Stream Video} & {\small Podcast Video} & {\small Embedded Music} & {\small Stream Audio} & {\small Stream Audio Leavebehind} & {\small Podcast} & {\small Podcast Leavebehind} \\
        \midrule
        Wide\&Deep  & 0\%  & 0\%  & 0\%  & 0\%  & 0\%  & 0\%  & 0\%  \\
        DCNv2 (t=1)  & 13.91\% & 21.39\% & 26.88\% & 15.91\% & 28.99\% & 19.11\% & \textbf{40.60\%} \\
        MMoE (t=2) & 3.34\%   & -0.64\%  & -35.68\%  & 2.34\%   & -12.74\% & 15.42\%  & -22.79\% \\
        CAMoE (t=2)  & \textbf{20.73\%} & \textbf{43.27\%} & \textbf{199.29\%} & \textbf{24.10\%} & \textbf{45.59\%} & \textbf{54.21\%} & 39.93\% \\
        CAMoE (t=7)  & 11.05\% & 24.60\% & 88.03\% & 9.17\% & 13.45\% & 27.00\% & 40.06\% \\
        \bottomrule
    \end{tabular}
\end{table*}

\begin{table*}[t]
    \centering
    \caption{AUC-PR comparison of a 2-task CAMoE model with content-based versus modality-based grouping (p-value < 0.05).}
    \label{tab:table3}
    \begin{tabular}{lcccc|ccc}
        \toprule
        \multicolumn{1}{c}{} & \multicolumn{4}{c|}{\textbf{\% change in AUC-PR for Music Slots}} & \multicolumn{3}{c}{\textbf{\% change in AUC-PR for 
 Podcast Slots}} \\
        \cmidrule(lr){2-5} \cmidrule(lr){6-8}
        & {\small Stream Video} & {\small Embedded Music} & {\small Stream Audio} & {\small Stream Audio Leavebehind} & {\small Podcast Video} & {\small Podcast} & {\small Podcast Leavebehind} \\
        \midrule
        \multicolumn{8}{l}{\textbf{Task Grouping}} \\
        Audio-Video    & \textbf{17.69\%}  & \textbf{227.98\%}  & \textbf{18.78\%}  & \textbf{29.82\%}  & \textbf{32.75\%}  & 37.10\%  & 5.55\% \\
        Music-Podcast & 4.69\%  & 123.60\%  & 1.81\%   & 14.99\%   & 23.38\%  & \textbf{40.58\%}  & \textbf{19.24\%}  \\
        \bottomrule
    \end{tabular}
\end{table*}

\subsubsection{Impact of Adaptive Loss Masking}

Table \ref{tab:table4} presents the effect of adaptive loss masking (ALM) on AUC-PR and Expected Calibration Error (ECE). In most cases, applying ALM leads to significant AUC-PR improvements for both audio and video, suggesting that allowing the model to focus solely on relevant labels enhances predictive performance. Without masking, video slots experience a notable AUC-PR decline, likely due to the model struggling to learn from irrelevant labels given the scarcity of video training data.

\begin{figure}[h]
    \centering
    \includegraphics[scale=0.28]{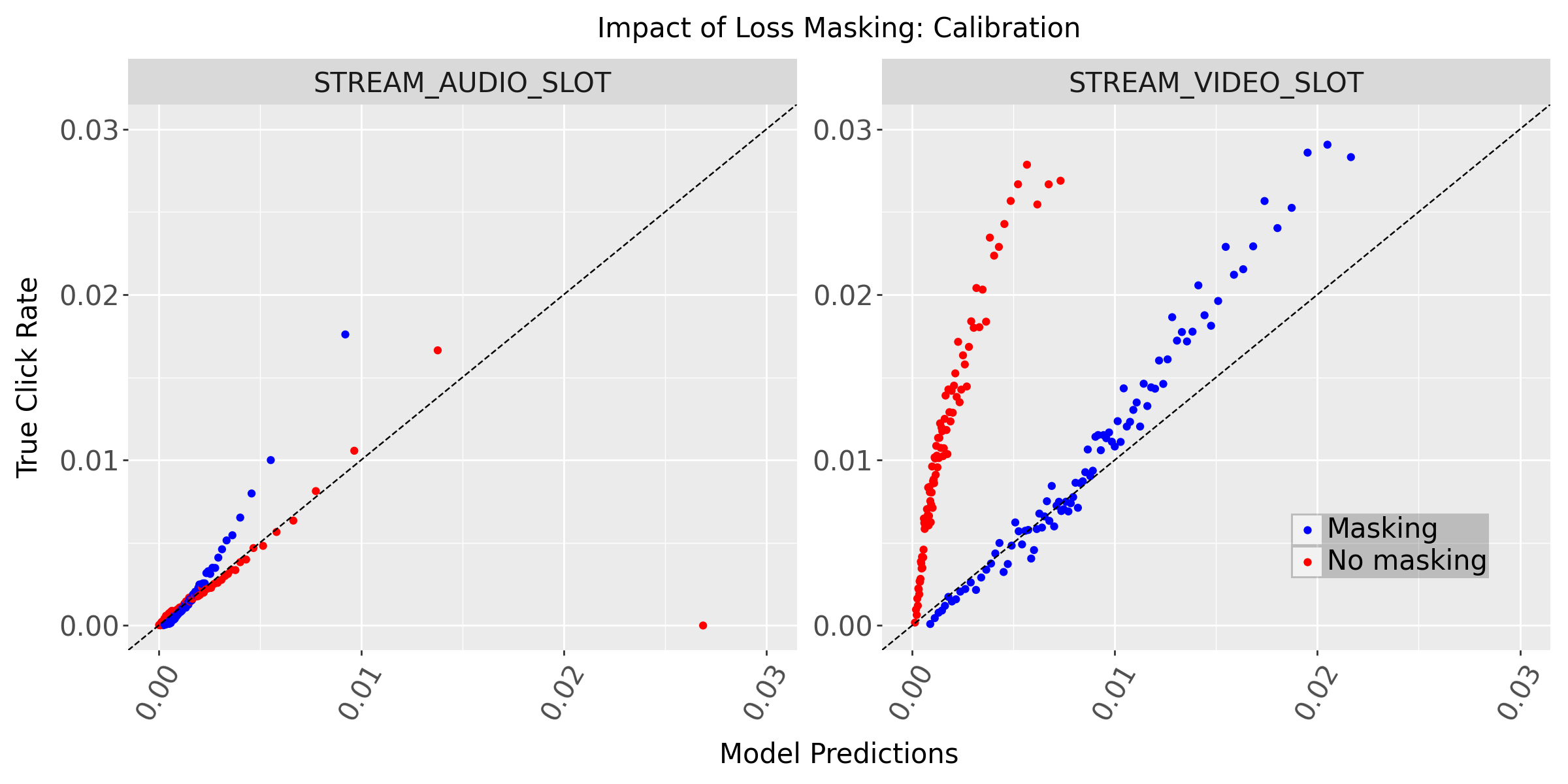}
    \caption{Calibration plots for the Stream Audio and Stream Video slots. Masking helps improve calibration for all slots except for Stream Audio.}
    \label{fig:masking_calib}
\end{figure}

Loss masking also improves ECE across most ad slots, particularly for \textit{Podcast Video}, reinforcing its role in enhancing calibration. By excluding irrelevant labels, the model better aligns predicted probabilities with observed outcomes. In contrast, without loss masking, \textit{Stream Video} exhibits the highest ECE, indicating poor calibration. As illustrated in Figure \ref{fig:masking_calib}, loss masking drastically improves calibration for \textit{Stream Video} calibration, while slightly worsening it for \textit{Stream Audio}.

These findings demonstrate that adaptive loss masking promotes task-specific learning by ensuring the model prioritizes relevant data. This effect is particularly pronounced for video-related slots, where removing loss masking allows audio patterns to dominate, making it harder for the model to learn video-specific interactions.

\begin{table*}[t]
    \centering
    \caption{Effect of adaptive loss masking (ALM) on AUC-PR and ECE, relative to the Wide\&Deep baseline (p-value < 0.05).}
    \label{tab:table4}
    \begin{tabular}{lccc|cccc}
        \toprule
        \multicolumn{1}{c}{} & \multicolumn{3}{c|}{\textbf{Video Slots}} & \multicolumn{4}{c}{\textbf{Audio Slots}} \\
        \cmidrule(lr){2-4} \cmidrule(lr){5-8}
        & {\small Stream Video} & {\small Podcast Video} & {\small Embedded Music} & {\small Stream Audio} & {\small Stream Audio Leavebehind} & {\small Podcast} & {\small Podcast Leavebehind} \\
        \midrule
        \multicolumn{8}{l}{\textbf{AUC-PR}} \\

        Masking    & \textbf{20.73\%} & \textbf{43.27\%} & \textbf{199.29\%} & \textbf{24.10\%} & \textbf{45.59\%} & \textbf{54.21\%} & \textbf{39.93\%} \\
        
        No Masking & -2.40\%  & -5.53\%  & -19.33\%  & 8.63\%   & 9.27\%   & 27.46\%  & 25.23\%  \\
        \midrule
        \multicolumn{8}{l}{\textbf{ECE}} \\
        Masking    & \textbf{-54.56\%}  & \textbf{-67.21\%} & -37.98\%  & -59.25\%  & \textbf{-56.95\%}  & \textbf{-66.32\%}  & \textbf{-49.51\%}  \\
        No Masking & 74.47\%   & -29.88\%  & \textbf{-60.82\%}  & \textbf{-69.35\%}  & -50.75\%  & -60.30\%  & -48.41\%  \\
        \bottomrule
    \end{tabular}
\end{table*}

\subsubsection{Impact of DCN} \label{sec:dcnimpact}
A key feature of our approach is the integration of Deep \& Cross Networks (DCN) within each expert of the mixture-of-experts (MoE) architecture. Table \ref{tab:table5} evaluates the effect of DCN on both MTL and single-task setups.

In the 2-task CAMoE setup, DCN provides substantial performance gains across all ad slots, with \textit{Embedded Music} improving the most, and even low-occurrence slots like \textit{Podcast Video} showing significant lifts. Without DCN, slot-wise performance is inconsistent — \textit{Stream Audio} and \textit{Stream Video} show minor gains, while \textit{Embedded Music} drops by 35.68\%, indicating that DCN plays a crucial role in modeling complex feature dependencies within CAMoE.

For the single-task model (Fig. \ref{fig:mtlabc}B), DCN improves performance, but less dramatically, as the lack of modality separation limits leveraging task-specific feature interactions. Interestingly, without DCN experts, the 2-task CAMoE performs worse than the the respective single-task setup, particularly for \textit{Stream Audio}, suggesting that DCNs facilitate effective transfer learning in multi-task settings by mitigating task interference and improving information sharing across modalities. Conversely, in a single-task setting, where feature interactions are inherently less complex, the added architectural complexity of DCN does not always translate to improved performance. This suggests that reducing task interference may be more beneficial than increasing model complexity. A uni-modal setup (Fig. \ref{fig:mtlabc}A) is also compared, exhibiting optimal Stream Video performance.

\begin{figure}[h]
    \centering
    \includegraphics[scale=0.29]{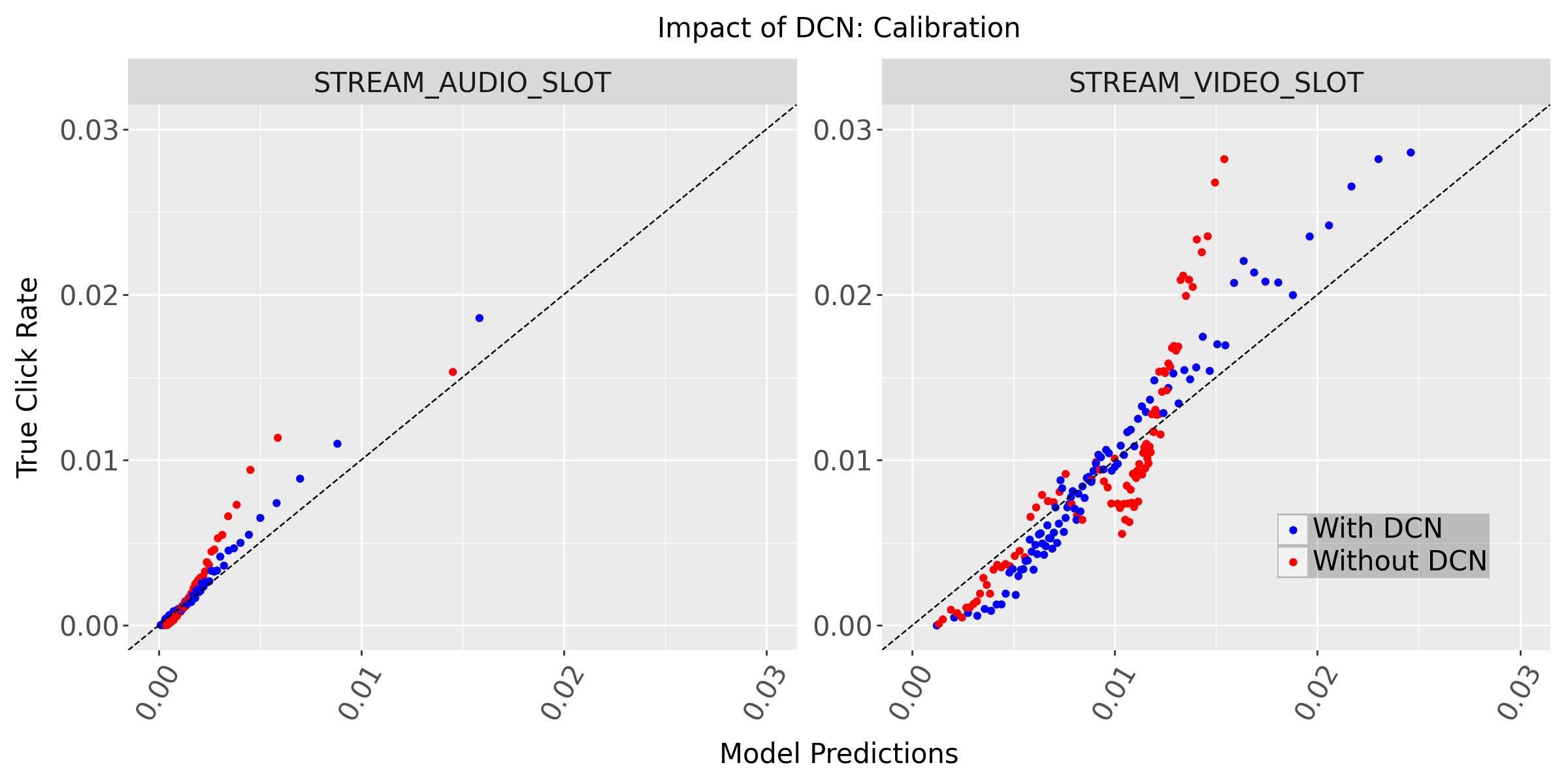}
    \caption{Calibration plots for the Stream Audio and Stream Video slots. Using DCN helps improve calibration for all slots.}
    \label{fig:dcn_calib}
\end{figure}

Fig. \ref{fig:dcn_calib} further illustrates the calibration improvements from DCN in the 2-task model, particularly for \textit{Stream Audio} and \textit{Stream Video}. Across all slots, DCN enhances calibration by reducing reliance on implicit feature interactions, preventing overfitting to noise, and lowering overconfidence in predictions. By acting as a stabilizer, DCN ensures smoother probability distributions, mitigating unpredictable shifts in click likelihood estimation.

\begin{table*}[t]
    \centering
    \caption{\% Effect of Deep \& Cross Network (DCN) as experts in the CAMoE framework, compared with uni-modal and single-task settings. Results are \% change in AUC-PR relative to the baseline. Results are statistically significant (p-value < 0.05).}
    \label{tab:table5}
    \begin{tabular}{lccc|cccc}
        \toprule
        \textbf{AUC-PR} & \multicolumn{3}{c|}{\textbf{Video Slots}} & \multicolumn{4}{c}{\textbf{Audio Slots}} \\
        \cmidrule(lr){2-4} \cmidrule(lr){5-8}
        & {\small Stream Video} & {\small Podcast Video} & {\small Embedded Music} & {\small Stream Audio} & {\small Stream Audio Leavebehind} & {\small Podcast} & {\small Podcast Leavebehind} \\
        \midrule
        \multicolumn{8}{l}{\textbf{2-task CAMoE model}} \\
        With DCN    & \textbf{20.73\%} & \textbf{43.27\%} & \textbf{199.29\%} & \textbf{24.10\%} & \textbf{45.59\%} & \textbf{54.21\%} & 39.93\% \\
        Without DCN & 3.34\%   & -0.64\%  & -35.68\%  & 2.34\%   & -12.74\% & 15.42\%  & -22.79\% \\
        \midrule
        \multicolumn{8}{l}{\textbf{Multi-modal single-task model}} \\
        With DCN    & 13.91\%  & 21.39\%  & 26.88\%  & 15.91\%  & 28.99\%  & 19.11\%  & 40.60\%  \\
        Without DCN & 7.93\%   & 9.31\%  & -12.39\%  & 9.93\%   & 5.71\% & 40.50\%  & \textbf{58.76\%} \\
        \midrule
        \multicolumn{8}{l}{\textbf{Uni-modal single-task model (Stream Audio or Stream Video Only)}} \\
        With DCN    & \textbf{22.75}\% & -  & -  & 11.18\%  & -  & -  & -  \\
        Without DCN & 10.30\%  & - & -  &   -42.33\% & - & - & - \\
        \bottomrule
    \end{tabular}
\end{table*}

\subsection{Online A/B Test Results}
For online evaluation, we measured effective cost-per-click (eCPC), CTR, and the number of clicks. A budget-split A/B test across 20 regional markets showed that the 2-task CAMoE model consistently outperformed the baseline across all these metrics, with statistically significant CTR improvements (see Table \ref{tab:table6}): 10\% for audio and 1.4\% for video slots. Post-deployment, CAMoE continued to drive measurable impact, with audio slots showing a 4.8\% eCPC decrease and a 14.5\% CTR increase, and video slots showing a 2.6\% eCPC decrease and a 1.3\% CTR increase. These results confirm the effectiveness of our multi-task approach in optimizing ad performance within audio-centric and multi-modal ad platforms.

\begin{table}[t]
    \centering
    \caption{A/B Test and 100\% Rollout Results (p-value < 0.005)}
    \label{tab:table6}
    \begin{tabular}{|l|c|c|c||c|c|}
        \hline
        & \multicolumn{3}{c||}{A/B Test Results} & \multicolumn{2}{c|}{100\% Rollout Results} \\
        \hline
        \textbf{Ad Slots} & \textbf{\# Clicks} & \textbf{eCPC} & \textbf{CTR} & \textbf{eCPC} & \textbf{CTR} \\
        \hline
        \textbf{Audio} & +12\% & -8\%  & +10\% & -4.8\% & +14.5\% \\
        \hline
        \textbf{Video} & +5\% & -1.2\%  & +1.4\% & -2.6\% & +1.3\% \\
        \hline
    \end{tabular}
\end{table}

\section{Conclusion} \label{conclusion}

In this work, we propose Cross-modal Adaptive Mixture-of-Experts (CAMoE) for multi-modal recommender systems and apply it to Spotify’s industrial-scale ad serving system for click-through rate (CTR) prediction. In an online A/B test, CAMoE improved CTR by 14.5\% for audio ads and 1.3\% for video ads. Offline experiments across various task groupings revealed that modality-based grouping is Pareto-optimal compared to content type, ad slot type, or single-task settings. We also demonstrated the importance of feature interaction in transfer learning. Given the critical role of calibration in ad serving, we further analyzed the impact of DCN experts on reducing calibration error in a multi-task model.

\begin{acks}
We would like to extend our thanks to the entire Ad Engagement team for their feedback and support: Kieran Stanley, Bharath Rengarajan, Nick Topping, Santiago Cassalett, Sanika Phatak, Ipsita Prakash, Gordy Haupt, Sneha Kadetotad, Ju Yang and Shaorong Yan.
\end{acks}

\vfill
\eject

\bibliographystyle{ACM-Reference-Format}
\balance
\bibliography{mtl-bib}

\appendix

\vfill
\eject

\section{Appendix}

\subsection{Pareto Optimality Analysis} \label{appendix:pareto}

In multi-objective optimization, a solution is considered \emph{Pareto optimal} if it's impossible to improve one objective without worsening at least one other objective. A set of Pareto optimal solutions forms the \emph{Pareto front}. In the context of our multi-task learning problem, we aim to optimize performance across multiple ad slots and metrics, making Pareto optimality a relevant concept for evaluating different model configurations.\\

\begin{figure}[h]
    \centering
    \includegraphics[scale=0.5]{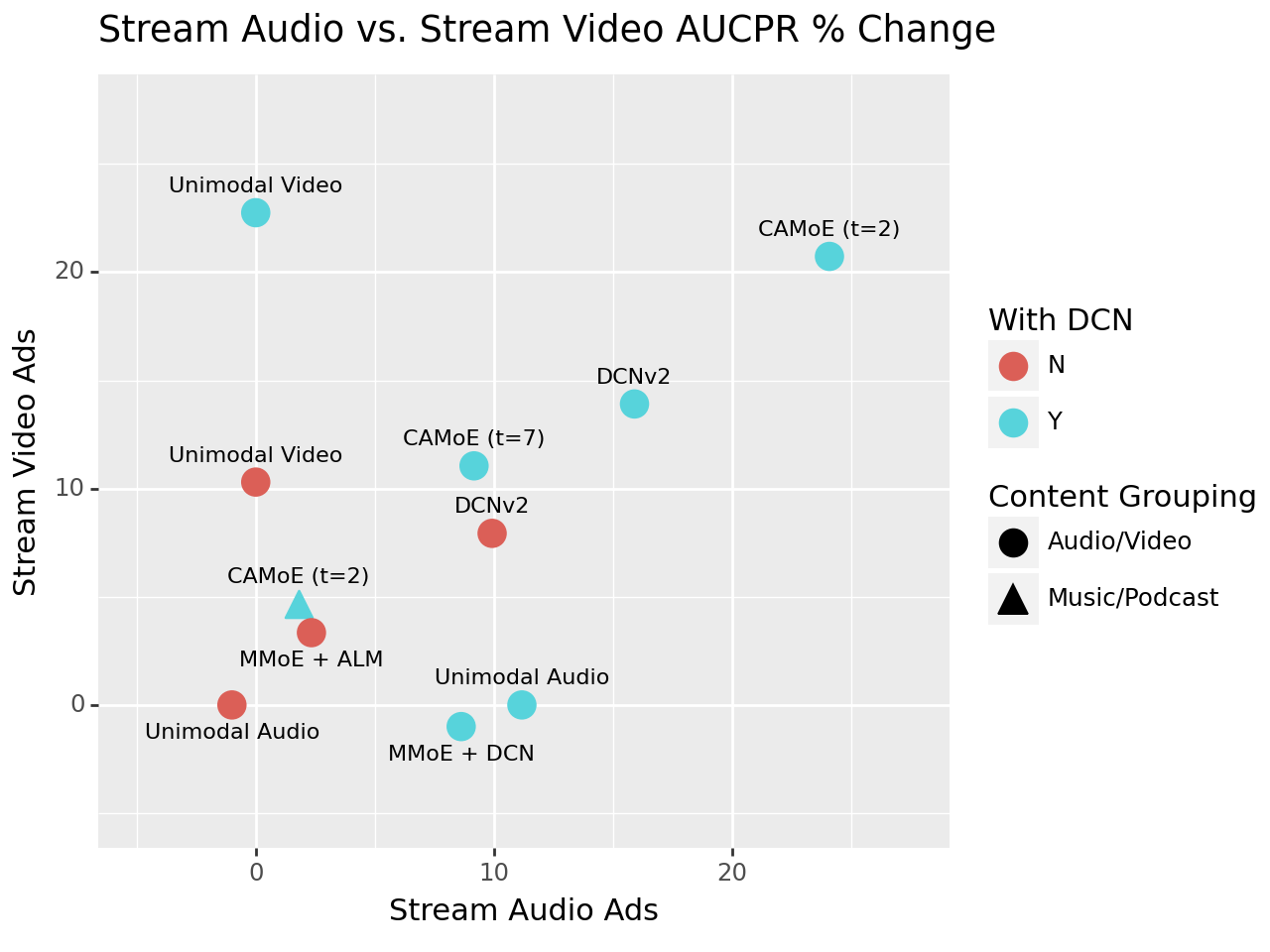}
    \caption{Analysis of Pareto Optimality for various CAMoE, multi-task and single-task configurations.}
    \label{fig:pareto_plot}
\end{figure}

Fig. \ref{fig:pareto_plot} presents a Pareto analysis comparing the performance of various models and configurations, focusing on the percentage change in AUC-PR for \emph{Stream Audio} and \emph{Stream Video} ads (relative to the Wide \& Deep baseline). These two slots are chosen as axes because they represent the two largest and highest-priority ad formats on Spotify. However, it's important to remember that our overall optimization considers performance across \emph{all} ad slots (as outlined in Table \ref{tab:slot_distribution}) and includes other objectives like calibration (ECE) and online metrics (CTR, eCPC).\\

Looking at Fig. \ref{fig:pareto_plot}:
\begin{itemize}
    \item The CAMoE (t=2) with DCN point in the top right of the graph is near Pareto Optimal. It dominates every point except for Unimodal Video.
    \item Unimodal models are not trained across multiple slot types, therefore are inferior for our use case.
\end{itemize}

\subsubsection{Broader Pareto Optimality Considerations}

While Fig. \ref{fig:pareto_plot} focuses on \emph{Stream Audio} and \emph{Stream Video}, a comprehensive Pareto analysis considers \emph{all} relevant objectives. Based on the data presented throughout the paper, we can make the following observations:

\begin{itemize}
    \item \textbf{Overall AUC-PR (Table \ref{tab:mtl_comparison}):} The 2-task CAMOE model achieves the highest average AUC-PR improvement across all ad slots compared to all other \emph{multi-task} models. While some individual slots might show slightly better performance with other configurations (e.g., DCNv2 for \emph{Podcast Leavebehind}), the 2-task model provides the best \emph{overall} balance.
    \item \textbf{Calibration (ECE):} The paper emphasizes the importance of calibration. While not explicitly plotted, the results in Subsection \ref{sec:dcnimpact} and Fig. \ref{fig:masking_calib} \& \ref{fig:dcn_calib}, along with the discussion of temperature scaling, suggest that the 2-task CAMOE model, particularly with adaptive loss masking and DCN, achieves good calibration across most slots. The 1-task model, while showing some AUC-PR gains, exhibits worse calibration.
    \item \textbf{Online Metrics (Table \ref{tab:table6}):} The A/B test results (Table 6) demonstrate that the 2-task CAMOE model significantly improves key online metrics (CTR and eCPC) compared to the baseline. This provides strong evidence of real-world Pareto optimality, as it improves performance without sacrificing other important business objectives.
    \item \textbf{Model Complexity and Maintainability:} While not explicitly quantified as an objective, the paper notes that the 7-task model has higher complexity and operational overhead. The 2-task model offers a good balance between performance and maintainability.
\end{itemize}

\subsubsection{Conclusion}

Based on the analysis of Fig. \ref{fig:pareto_plot}, combined with the broader results presented in Tables \ref{tab:mtl_comparison}-\ref{tab:table6}, we conclude that the \emph{2-task CAMoE model with DCN and adaptive loss masking} represents a near-Pareto optimal solution for the multi-modal ad targeting setup on Spotify. It achieves the best overall balance between AUC-PR across key ad slots, calibration, online performance (CTR and eCPC), and model complexity. While other configurations might show slight advantages for \emph{individual} objectives, the 2-task CAMOE model provides the most effective and practical solution for the overall system.

\subsection{Expert Specialization Study} \label{appendix:expertstudy}

To further investigate the specialization of experts within our 2-task CAMoE model and the potential impact of data imbalance, we conducted an experiment involving \textit{inference-time expert masking}. This technique is distinct from standard dropout, which is applied during training. Here, we selectively mask the entire output of an expert during inference, forcing the model to rely on the remaining expert(s). This study explores the degree to which each expert has learned modality-specific representations.

\subsubsection{Problem and Motivation}

Despite the modality-aware task grouping and Adaptive Loss Masking (ALM) employed in CAMoE, some degree of implicit influence between the audio and video experts can still occur due to the shared embedding layer and gating network (see Section \ref{architecture}). This influence, stemming from the dominant audio data, could potentially hinder the video expert's specialization and limit performance on video ad CTR prediction. We hypothesized that selectively masking experts at inference could mitigate this influence and improve video performance.

\subsubsection{Experimental Setup}

We used the trained 2-task CAMOE model as the basis for this experiment. We did not retrain the model. During inference, we evaluated three configurations:

\begin{enumerate}
    \item \textbf{Left Mask:} The output of Expert 1 (arbitrarily designated as the "left" expert) is set to zero before being passed to the task-specific towers. This forces the model to rely solely on Expert 2 for predictions.
    \item \textbf{Right Mask:} The output of Expert 2 (the "right" expert) is set to zero, forcing reliance on Expert 1.
    \item \textbf{No Mask:} The standard inference procedure, where both experts contribute (this is our baseline CAMoE model, as deployed in production).
\end{enumerate}

We evaluated performance on the held-out test set, focusing on the relative change in AUC-PR for the \textit{Stream Video} ad slot. We chose this slot because CAMoE's improvements, as described in Section 6, were most pronounced for \textit{Stream Audio} ads, making it a sensitive indicator of expert specialization.

\subsubsection{Modified Inference with Expert Masking}
Let $E_k(x)$ represent the output of expert $k$ given input $x$, and $G_m(x)$ be the output of gate for task $m$. Let $M_k$ be the mask for the expert, we define it as follows:

\begin{equation}
M_k =
\begin{cases}
    0, & \text{if expert } k \text{ is masked} \\
    1, & \text{otherwise}
\end{cases}
\end{equation}

Then, recall from the paper that we apply a softmax on the embedding layer via task-specific gates:
\begin{equation}
    \text{GateOutput}_{m}(x) = \text{Softmax}(G_m(x))
\end{equation}

Then, the modified output, $\hat{y}^{(m)}$, for task $m$ with expert masking is:
\begin{equation}
\hat{y}^{(m)}(x) = \sigma\left(T_m\left(\sum_{k=1}^{K}  \text{GateOutput}_{m}(x)_k  \cdot (M_k \cdot E_k(x)) \right)\right)
\end{equation}

where:
\begin{itemize}
\item $K$ is the total number of experts (in our case, $K=2$).
\item $\text{GateOutput}_{m}(x)_k$ is the $k$-th element of the softmax output from the gating network for task $m$. This represents the weight assigned to expert $k$ for task $m$.
    \item $E_k(x)$ is the output of expert $k$.
    \item $M_k \cdot E_k(x)$ applies the mask: if $M_k=0$, the expert's output is zeroed out.
    \item $T_m$ is the task-specific tower for task $m$.
    \item $\sigma$ is the sigmoid function.
\end{itemize}

\subsubsection{Results}

The results for \textit{Stream Video} AUC-PR change (relative to the 2-task CAMoE performance as per Table \ref{tab:mtl_comparison}) are reported in Table \ref{tab:table7}.

\begin{table}[t]
    \centering
    \caption{Effect of inference-time expert masking on AUC-PR of Video ad slots, relative to the 2-task CAMoE model (\textit{No Mask}) in Table 2 (p-value < 0.05).}
    \label{tab:table7}
    \begin{tabular}{lccc}
        \toprule
        {\textbf{Mask}} & {\small Stream Video} & {\small Podcast Video} & {\small Embedded Music} \\
        \midrule

        None & 0\%  & 0\%  & 0\%  \\
        Left Expert   & 6.58\% & 1.62\% & 67.99\% \\
        Right Expert  & 6.47\% & -3.01\% & 122.4\% \\
        
        \bottomrule
    \end{tabular}
\end{table}

\subsubsection{Analysis and Discussion}

The results demonstrate that inference-time expert masking has a variable impact depending on the specific video ad slot and which expert is masked. This reinforces the idea of asymmetrical specialization, but also reveals a more complex interplay of factors than initially hypothesized.

\textbf{Stream Video.}  Both left and right masking significantly improve Stream Video AUC-PR (by approximately 6.5\%). This strongly supports the idea that both experts contain some information relevant to \textit{Stream Video}, but that masking one expert removes interfering signals (either from the shared components or from the other expert's less robust video features), leading to better performance. This confirms that the model has learned a degree of specialization.\\
    \textbf{Podcast Video.}  Left masking (using only Expert 2) provides a small improvement (+1.62\%), while right masking (using only Expert 1) leads to a decrease in performance (-3.01\%). This suggests that Expert 2 has learned representations that are more beneficial for \textit{Podcast Video} than Expert 1.  Expert 1 might contain some features that are detrimental to \textit{Podcast Video} prediction, perhaps due to overfitting to the more abundant Stream Video or general audio data.\\
    \textbf{Embedded Music.}  Left masking (using only Expert 2) yields a large improvement (+67.99\%), while right masking (using only Expert 1) provides an even more substantial improvement (+122.40\%). This indicates that Expert 1 is \textit{significantly} more specialized in features relevant to \textit{Embedded Music}.  Expert 2 also contains useful information, but to a lesser extent. The magnitude of these improvements suggests that \textit{Embedded Music}, likely due to its relative data scarcity compared to \textit{Stream Video}, is particularly sensitive to the benefits of expert specialization and the removal of interfering signals.\\
     \textbf{Shared and Specialized Representations.}  As hypothesized, both experts appear to learn a mix of general and modality-specific features.  The data imbalance and shared model components (embedding layer, gating network) likely lead to asymmetrical specialization, with one expert becoming \emph{more} (but not \emph{exclusively}) focused on audio and the other \emph{more} focused on video. The varying impact across video slots suggests that the degree of specialization, and the specific features learned by each expert, differ depending on the characteristics of each slot.\\
    \textbf{Noise Reduction.} This experiment confirms that noisy activations from one of the experts was impacting the performance of the \textit{Stream Video} slot.

\subsubsection{Deployment Considerations}

While the expert dropout experiment demonstrates significant potential for performance gains, particularly for \textit{Embedded Music} and \textit{Stream Video}, \emph{we did not deploy this modification to production}. The primary reason is the potential for increased inference latency. A naive implementation that evaluates both masked configurations would double the computation time, which is unacceptable in our low-latency ad serving environment. Future work will focus on addressing these serving challenges and exploring more sophisticated masking strategies.

\end{document}